\documentclass[aps,prd,preprint,groupedaddress,nofootinbib,tightenlines]{revtex4}
\usepackage{epsfig}
\usepackage{latexsym}

\newcommand{\beq}{\begin{eqnarray}}
\newcommand{\eeq}{\end{eqnarray}}

\newcommand{\pbp}{\langle \bar{\psi} \psi \rangle}

\begin{document}

\title {QCD thermodynamics from an Imaginary $\mu_B$: \\
results on the four flavor lattice model} 

\author{Massimo D'Elia}
\email{delia@ge.infn.it}
\affiliation{Dipartimento di Fisica dell'Universit\`a 
di Genova and INFN, I-16146, Genova, Italy }
\author{Maria-Paola Lombardo}
 \email { lombardo@lnf.infn.it}
\affiliation {INFN-Laboratori Nazionali di Frascati, 
I-00044, Frascati(RM), Italy}

\begin{abstract}
We  study four flavor QCD at nonzero temperature and density
by analytic continuation from an imaginary chemical potential.
The explored region is $T = 0.95 T_c < T < 3.5 T_c$, and the
baryochemical potentials range from 0 to $\simeq 500$ MeV.
Observables include the number density,  the order parameter
for chiral symmetry, and the pressure, which 
is calculated via an integral method at fixed temperature and quark mass.
The simulations are carried out on a $16^3 \times 4$ lattice, 
and the mass dependence of the results is estimated by exploiting
the Maxwell relations. In the hadronic region we confirm that
the results are consistent with  a simple resonance hadron gas model,
and we estimate  the critical density by combining the results for the
number density with those for the critical line. In the hot phase, 
above the endpoint of the Roberge-Weiss transition $T_E \simeq 1.1 T_c$ 
the results are consistent with a free lattice 
model with a fixed effective
number of flavor slightly different from four.
We confirm that  confinement and chiral 
symmetry are coincident by a further analysis of the critical
line, and we discuss the interrelation between
thermodynamics and critical  behavior. We comment on the
strength and weakness of the method, and propose further developments.
\vspace{1pc}
\end{abstract}
\maketitle
\section{Introduction}

QCD at nonzero temperature and density is an important subject
both from a theoretical and phenomenological perspective\cite{KogutStephanov}.
Current and future experiments at RHIC and LHC will explore
the  region of the phase diagram close to the zero density axis,
not far from the cooling path of the primordial universe. The cold,
high density phases are relevant for astrophysics. Future GSI experiments
might bridge these two regimes and explore the region of intermediate
temperatures and densities. 

While many  predictions on the structure of the phase diagram
can  be obtained by use  of simple models dictated by symmetry arguments,
quantitative studies require a first principle calculation. Lattice
field theory is the natural approach, and QCD at nonzero temperature
and density is by now studied by a variety of lattice methods:
the multiparameter reweighting method achieves on optimal overlap 
between the simulation ensemble at zero baryon density
and the target ensemble at nonzero density
\cite{Fodor:2001au,Fodor:2002km,Fodor:2001pe,Fodor:2004nz}; 
the direct calculation of the 
derivatives gave the first informations on the physics at nonzero chemical
potential \cite{Gavai:2003mf, Gavai:2003nn, deForcrand:2002pa}; 
the Taylor expanded reweighting reduces the numerical costs
associated with the multiparameter 
reweighting \cite{Allton:2002zi, Allton:2003vx};
the analytic continuation from an imaginary
chemical potential uses the information from the $\mu^2 \le 0$ halfplane
to reconstruct the physics of real baryon density
\cite{Lombardo:1999cz,Hart:2000ef,
deForcrand:2002ci,D'Elia:2002gd,deForcrand:2003hx}.
In addition to these pragmatic works, which take advantage of the fluctuations
close to $T_c$ to explore the small chemical potential region $\mu/T \le 1.$,
there have been a number of  new proposals
\cite {Ambjorn:2002pz,Azcoiti:2003vv,Crompton:2001ws}, 
discussions\cite{Ejiri:2004yw} and 
checks\cite{Giudice:2004se}.
For recent reviews see 
\cite{Laermann:2003cv, Katz:2003up}, 
and for introductions into the 
subject see \cite {Muroya:2003qs,Lombardo:2004uy}.

In this paper we extend our study of four flavor QCD within the imaginary
chemical potential approach. Let us remind that  four flavor
QCD has a first order transition at $\mu=0$, as it is the one 
at $T=0$. Hence, it is expected to have a first order critical line 
in the plane $T,\mu$: there is no 
tricritical point or endpoint to be investigated here.
On the other hand, the first order (or sharp crossover)
nature of the critical line  makes the model simpler,  and amenable 
to a detailed study with comparatively modest numerical resources.

In our first paper \cite{D'Elia:2002gd} we studied
the order parameter in the hadronic phase and calculated the critical
line up to $\mu \simeq 500 MeV$. 
We have indeed confirmed the expected first order
(or very sharp crossover) nature of the transition,
we have observed its interrelation with deconfinement, and we have studied
its $N_f$ scaling by comparing  our results with those of ref.
\cite{deForcrand:2002ci}. The results on the chiral condensate
were consistent with a partition function described by
a simple hyperbolic cosine behavior. Consistent numerical
findings were reported
in ref. \cite{Karsch:2003zq}, and intepreted within a resonance gas model.

This work addresses more fully the properties of the hadronic phase,
and those of the plasma phase. We base our analysis mostly
on the results for the number density and for the chiral condensate.
We  calculate the subtracted pressure by an integral method at
constant temperature and quark mass, and we study the mass dependence
by taking numerical derivatives of the chiral condensate.

The properties of the quark gluon plasma 
phase are studied by means of the number density and
of the subtracted pressure.
We discuss the nonperturbative nature of the hot phase
in the vicinity of the critical point, and we argue that this
result follows naturally by an analysis of the phase diagram in the
$T, \mu^2$ plane. We  compare the numerical results with
analytic predictions, a task for which the imaginary chemical potential
approach is ideally suited: for instance, rather than analytically continue
the numerical results to real chemical potential, we can continue the
analytic predictions to the negative $\mu^2$ halfplane, and contrast
the results with the numerical ones.

In the hadronic phase we study in detail the number density: its
simple behavior will further support the applicability of the hadron resonance
gas model; moreover we  combine the results for
the critical line with those for the number density in order to
estimate the critical density.

In both phases we  exploit the Maxwell relations to study the
mass dependence: this  turns out to be sizable in the
hadronic phase, and negligible in the plasma phase.

Concerning the critical line, as already mentioned,
the four flavor model has
a first order transition. Hence, tricritical points or endpoints
are not present here, and we contented ourselves with the
precision reached in our previous study. The new results on the
critical behavior consists in a more detailed analysis of the
correlation of the chiral and the deconfinement transition.

The paper is organised as follows. In Section II we review
the properties of the phase diagram of QCD in the chemical potential--
temperature plane, including the new results on the critical behavior
at a selected $\mu$ value. 
In Section III we describe our observables and give an overview of the 
results. Section
IV is devoted to the results in the hadronic region,  while Section V
presents results for the Quark Gluon Plasma phase. In either phases
we will discuss in detail the various {\it ans\"atze}
which emerge naturally  once the analyticity
properties and the nature of the critical lines are taken into account.
In Section VI we discuss our
results for $T_c < T < T_E$ and give some general comment about this 
intermediate region of temperatures. We summarize and
discuss future perspectives in Section VII.
Some preliminary results have already appeared in 
\cite{D'Elia:2003uy}.

\section{QCD in the $\mu^2, T$ plane}

Results from simulations with an imaginary chemical potential
can be analytically continued to a real chemical potential,
thus circumventing the sign problem 
\cite{Hart:2000ef,deForcrand:2002ci,D'Elia:2002gd}. 
In practice, the analytical
continuation is carried out along one line in the complex
$\mu$ plane: first along the
imaginary axes, and then along the real one. It is then
meaningful to map  this path in the complex $\mu^2$ plane: because
of the symmetry property $Z(\mu)=Z(-\mu)$ this can be achieved 
without losing generality. In the complex $\mu^2$ plane
the partition function is real for real values of the external parameter
$\mu^2$, complex otherwise: the situation resembles
that of ordinary statistical models in an external field. Hence,
the analyticity of the physical observables \cite{Lombardo:1999cz} 
as well as that of the critical line 
\cite{deForcrand:2002ci} follows naturally.
 
The reality region for the partition function represents states which
are physically accessible. The reality region for the determinant
represents the region which is amenable to an importance
sampling calculation: $\Re e  \mu^2 \le 0$. 

\begin{figure}
{\epsfig{file= 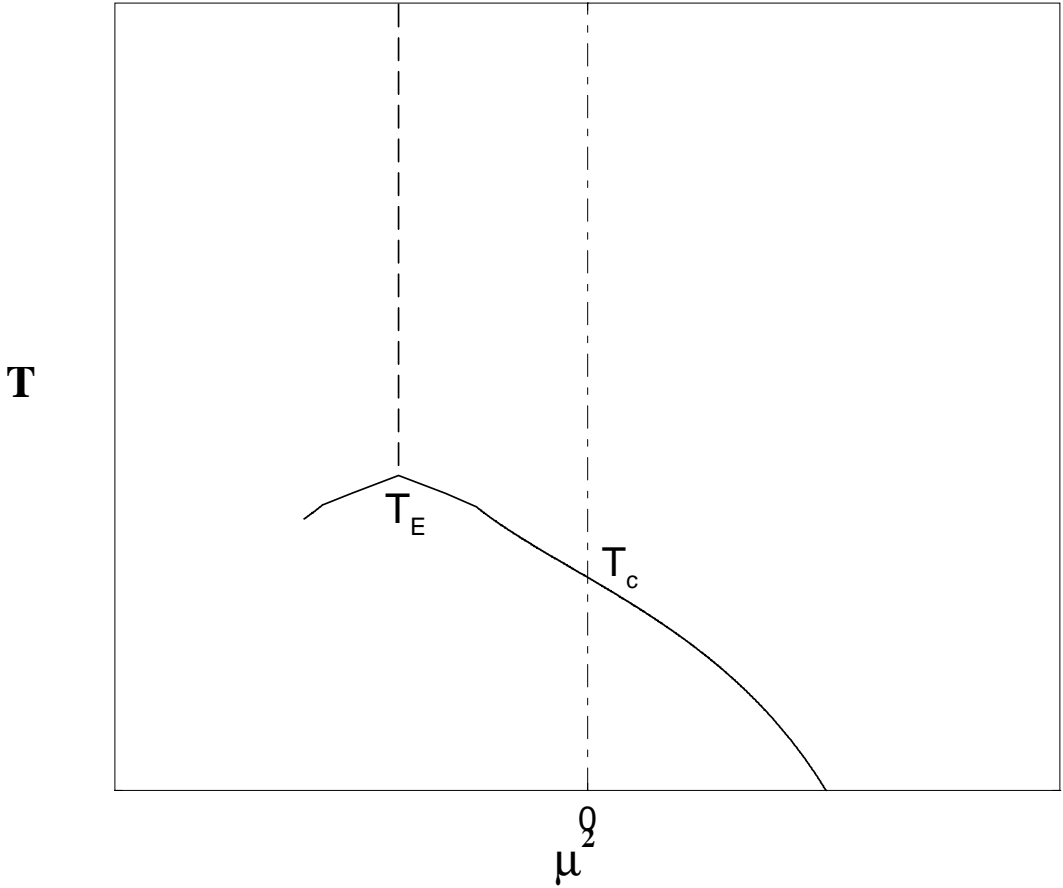, width= 14 truecm}}
\caption{Sketch of the phase diagram in the $\mu^2,T$ plane:
the solid line is the chiral transition, the dashed
line is the Roberge Weiss transition. Simulations can be carried out at
$\mu^2 \le 0$ and results continued to the
physical domain $\mu^2 \ge 0$.
The derivative and reweighting methods have been used so far
to extract informations from simulations performed at $\mu=0.$
The imaginary chemical potential approaches uses results on the
left hand half plane. Different methods could be combined 
to improve the overall performance}
\end{figure}

\begin{figure}[htb]
%\framebox[55mm]{\rule[-21mm]{0mm}{43mm}}
\psfig{file=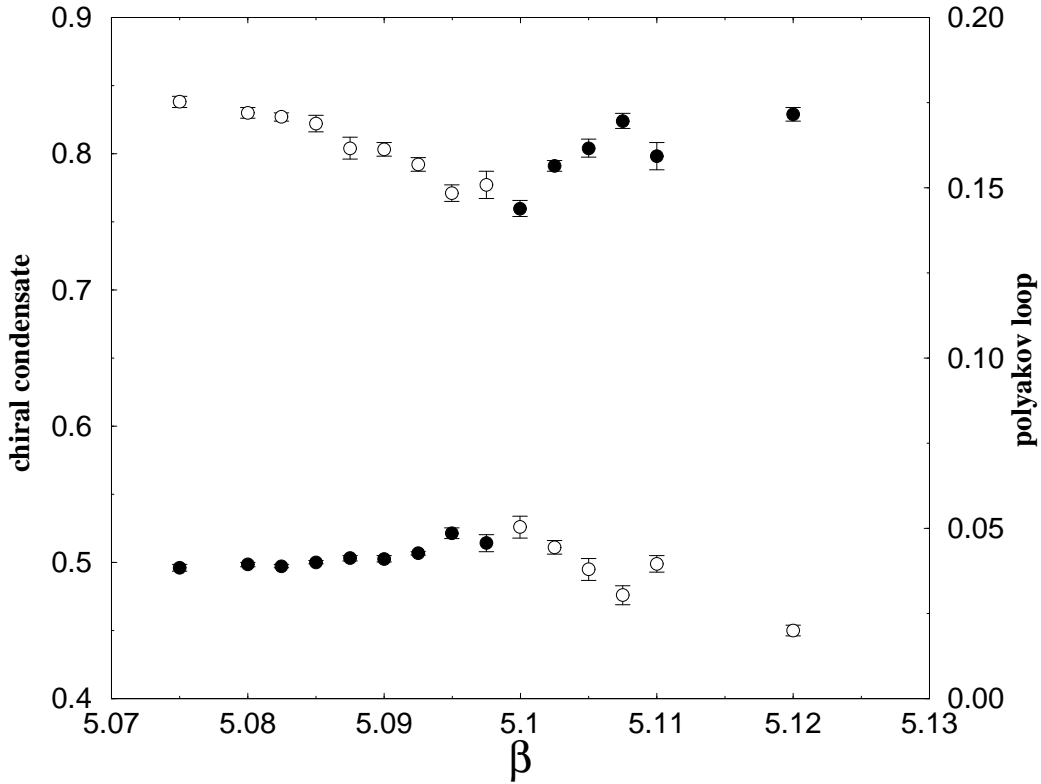,width=14 truecm}
\caption{Correlation between $\pbp$ and Polyakov loop
at $\mu_I = 0.15$, demonstrating the correlation of chiral
and deconfining transition at nonzero baryon density.}
\label{fig:pol_psi}
\end{figure}

\begin{figure}[htb]
%\framebox[55mm]{\rule[-21mm]{0mm}{43mm}}
\psfig{file=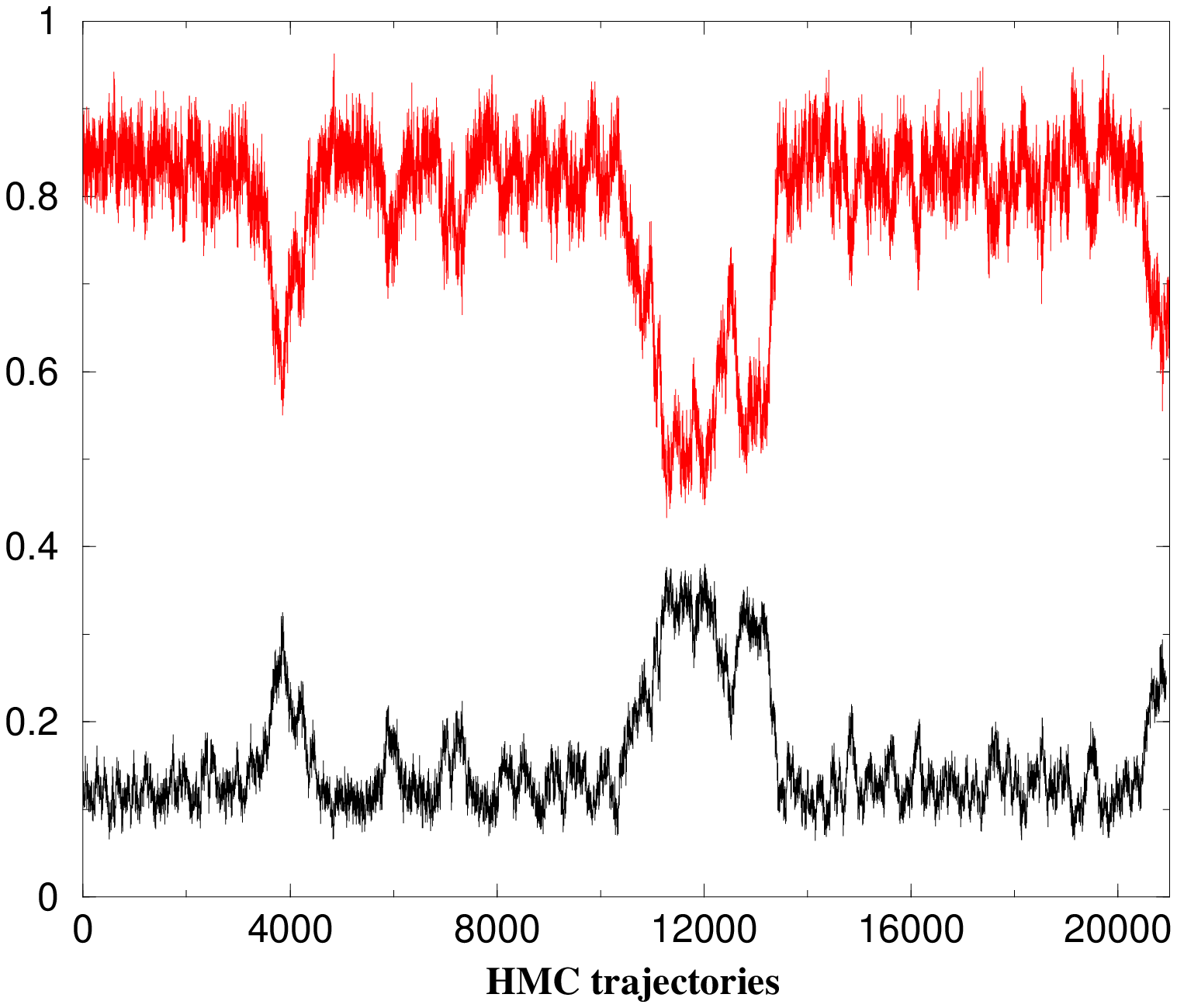,width=14 truecm}
\caption{Monte Carlo histories of the chiral condensate
(upper history) and of twice the Polyakov loop modulus (lower
history) for $\mu_I = 0.15$ at the phase transition. The apparent
correlation between the two quantities is the clearest demonstration
of the coincidence of deconfinement and chiral phase transition
also for $\mu_I \neq 0$.}
\label{fig:chiral_poly_p}
\end{figure}

The phase diagram in the temperature, (real) $\mu^2$ plane is
sketched in Fig.1, where we omit the superconducting and the color flavor
locked phase, which (unfortunately) play no r\^ole in our discussion.
The region accessible to numerical simulations is the one with
$\mu^2 \le 0$: at a variance with other approaches to finite density
QCD,   which so far only used information at $\mu=0$ 
\cite{Fodor:2001au,Allton:2003vx,Crompton:2001ws},
the imaginary chemical potential method exploits the entire halfspace.

Note that after the rotation to $\mu^2$ there is no analytic
continuation in complex plane to be done, but rather we can talk about
a simpler analytic extrapolation along the real $\mu^2$ axis.

We note here that there are physical questions which can 
be addressed quite simply: in Fig. \ref{fig:pol_psi}
we demonstrate the correlation between the average values of the chiral condensate and
of the Polyakov loop. In Fig. \ref{fig:chiral_poly_p} the same
correlation is illustrated directly on the Monte Carlo time histories at
the phase transition,
showing the striking result that not only the two phase transitions are
coincident, but that the chiral condensate ant the Polyakov loop
are completely correlated even at the level of small scale fluctuations. 
The data are obtained at fixed $\mu_I = 0.15$,
and a similar behavior can be observed for other values of $\mu_I$,
including zero \cite{Karsch:2001cy}.
This correlation should be continued at real baryon density:
to this end, we note that 
if $\beta_c(i \mu_I) = \beta_d(i \mu_I) $ over a finite
imaginary chemical potential interval, then the function 
$\Delta {\beta} (i \mu_I) = 
\beta_c(i \mu_I) - \beta_d(i \mu_I)$ is simply continued to be zero over
the entire analyticity domain, thus demonstrating the correlation
between the chiral and the deconfinement transitions ($\beta_c = \beta_d$)
also for real values of $\mu$. We refer to \cite{Mocsy:2003qw} for
an effective Lagrangian discussion of this issue, and to  \cite{Alles:2002st}
for analogous results in the two color model.

\section{Observables and Overview of the results}

The numerical simulations  were performed, using the HMC algorithm,
with four flavor of staggered fermions, on the same lattice 
$16^3 \times 4$ and with the same mass $am = 0.05$ as in our previous study. 
New numerical results have been obtained in the plasma phase, for
$\beta = 5.310, 5.650, 5.689$, corresponding
to $T \simeq (1.5 T_c, 2.5 T_c, 3.5 T_c)$ (we used the two loop
$\beta $ function to convert to physical units), and statistics 
for $ \beta = 5.030 (T \simeq 0.985 T_c)$ 
and $\beta = 5.100 (T \simeq 1.095 T_c)$, close to the Roberge Weiss
transition point, have been improved as well.

Our analysis is mostly based on measurements of the chiral condensate,
number density and  Polyakov loop. Note that the number
density is purely imaginary for imaginary chemical potential:
in the following $n(T, \mu_I, m_q)$  will denote 
the imaginary part of the result. 
From  $\pbp$ and $n$
 we will calculate and exploit  derivatives and integrals with respect to the
chemical potential, yielding the mass dependence of the number density,
the quark number susceptibility , and the pressure.

\begin{figure}[htb]
%\framebox[55mm]{\rule[-21mm]{0mm}{43mm}}
\psfig{file=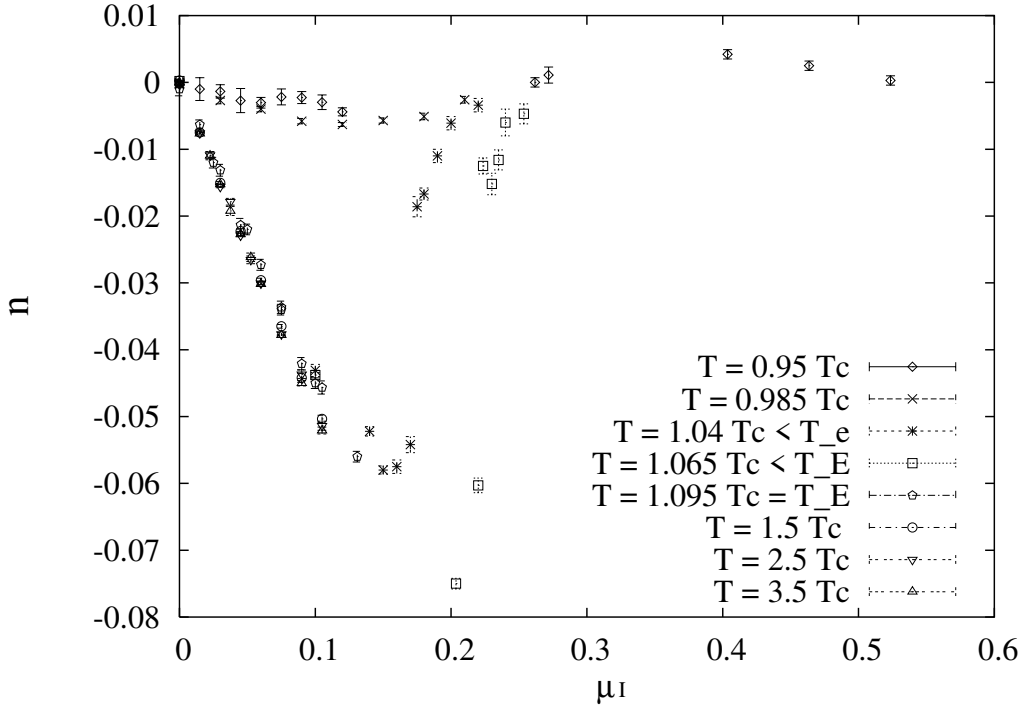,width=14 truecm}
\caption{Overview of the results for the baryon density
as a function of $\mu_I$: the behavior
is smooth in the hadronic phase, shows the expected discontinuity associated
with the chiral/deconfining transition in the intermediate region 
$T_c < T < T_E$, and increases rapidly in the quark gluon plasma phase. }
\label{fig:dens}
\end{figure}

In Figure \ref{fig:dens} we present an overview of our
results for the number density, where we can read off the main features 
outlined in the discussion of the phase diagram of Section II above: 
for $T < T_c$ the results are smooth; in the intermediate region 
$T_c < T < T_E$\footnote{We will use $T_E$
to denote the endpoint of the Roberge Weiss transition; as there is
no QCD endpoint in this model no confusion should arise}, 
there is a clear discontinuity, in correspondence with the
chiral/deconfining transition; finally, above $T_E$ , the results
for the number density increase sharply, eventually approaching the
free field results.

Slightly anticipating the numerical analysis which will be presented below,
we show in Figure \ref{fig:susc} the summary of our results for the quark
number susceptibility $\chi_q$:
\begin{equation}
\chi_q(T,\mu=0,m_q) = \frac {\partial n(T,\mu_I,m_q)}{\partial \mu}|_{\mu=0}
\label{eq:chi}
\end{equation}

Obviously, in the case of a polynomial behavior
$n(T,\mu_I,m_q) = a(T,m_q)\mu_I + b(T,m_q) \mu_I^3 $, 
$\chi_q(T,\mu=0,m_q)= a (T, m_q)$. 
For other fits $\chi_q(T,\mu=0,m_q)$  is
obtained from equation (\ref{eq:chi}), taking for $n(T,\mu_I,m_q)$ 
the corresponding fitted function. 
The agreement between different analysis can be judged from Figure
\ref{fig:susc}.

One word of comment might be in order: aside from the free massless 
case, the analytic form for $n(T,\mu_I,m_q)$ is, of course,  
not known. Given that the numerical results have errors,
different ans\"atze might well produce equally satisfactory results (measured,
for instance, by the $\chi^2/d.o.f$). When this is the case, we have checked
that the analytically continued results from the two different forms are
consistent with each other, within errors. The results presented in
 Figure \ref{fig:susc} offer one first example of this check.

To calculate the pressure, or rather, its
variation with $\mu$ 
\footnote{From now on $\mu$ will denote the real
chemical potential, i.e. the complex chemical potential is
$\mu_{complex} = \mu + i \mu_I$} at constant  temperature and quark mass
\begin{equation}
\Delta P/T^4 = (P(T,\mu, m_q) - P(T, \mu=0, m_q))/T^4
\end{equation}
we exploited the relationship
\begin{equation}
n(T,\mu,m_q) = \frac{\partial P(T,\mu,m_q)}{\partial \mu}
\end{equation}
which gives
\begin{equation}
(P(T,\mu, m_q) - P(T, \mu=0, m_q))/T^4 = N_t^4 \int n(\mu) d\mu 
\end{equation}

This can be achieved by numerical  integration of the
results for the number density $n(T, \mu_I, m_q))$, thus obtaining
$\Delta P(T, \mu_I, m_q) /T^4$ (which can be continued to real
chemical potential). A similar approach to the calculation of
pressure can be pursued  at zero baryon 
number \cite{Kratochvila:2003rp}.

In Figure \ref{fig:pressI} we show the results of 
$\Delta P(T, \mu_I, m_q) /T^4$ obtained in this way. 
The results for the free case have been obtained by calculating 
$n$ on a $12^3 \times 4$ lattice, for $m_q = .05$, fitting the
resulting data to a third order polynomial, then continued
to imaginary chemical potential, which amounts to a flip of
the third order term.  Actually, for
$T > 1.1 T_c$ a linear term suffices to describe the data,
the third order term can be set to zero, and
the data in Figure  \ref{fig:pressI} coincide with their
analytic continuation to a real chemical potential in that
temperature range.

In Figure \ref{fig:pressII} we plot the same data in a different
form, for the sake of an easier comparison with  results from 
other groups \cite{Fodor:2002km,Allton:2003vx,Szabo:2003kg,Csikor:2004me}. 
We also note that an alternative procedure to 
obtain $\Delta P(T, \mu,  m_q)/T^4$ 
is of course by analytical integration of the
results for $n(\mu)$ obtained by analytic continuation:
the two procedures  give consistent results.

\begin{figure}[htb]
%\framebox[55mm]{\rule[-21mm]{0mm}{43mm}}
\psfig{file=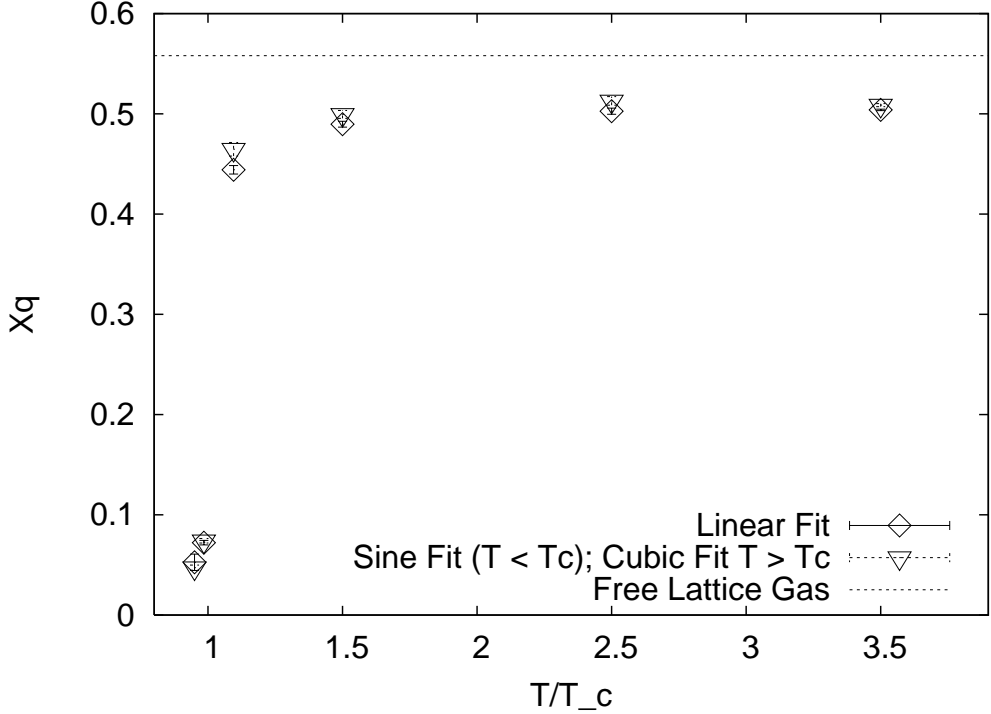,width=14 truecm}
\caption{Particle number susceptibility at $\mu=0$ as a 
function of temperature from different fits (see text for details).
The susceptibility is small, yet different from zero, in the hadronic
phase close to $T_c$, and 
reaches a nearly constant value already for $T \simeq 1.5 T_c$. }
\label{fig:susc}
\end{figure}

\begin{figure}[htb]
%\framebox[55mm]{\rule[-21mm]{0mm}{43mm}}
\psfig{file=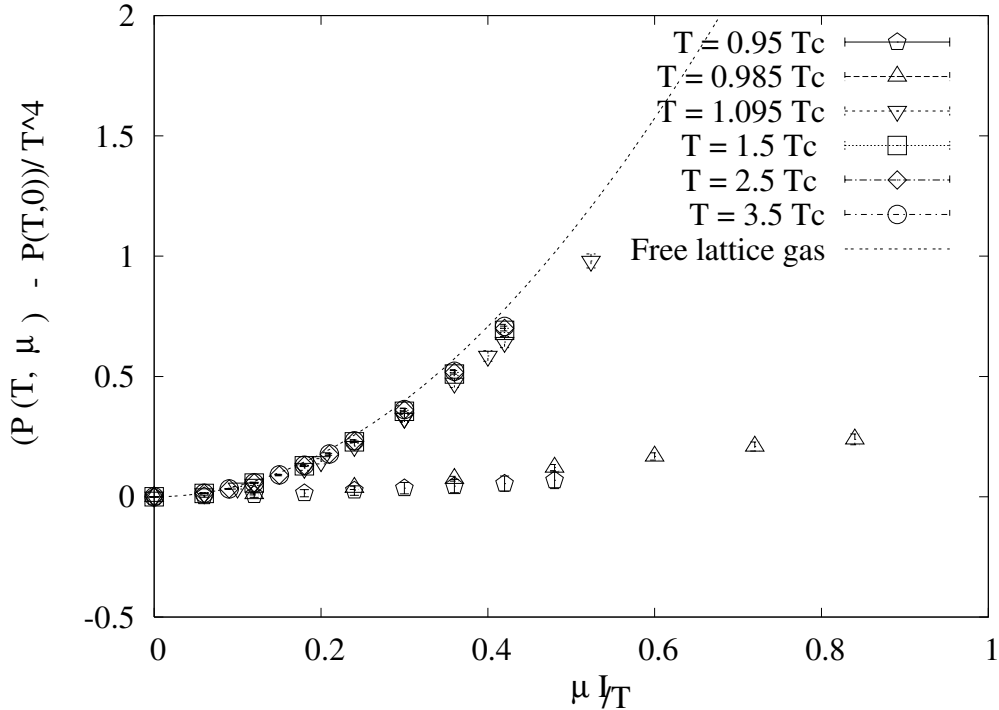,width=14 truecm}
\caption{Overview of the results for $\Delta P (T, \mu_I, m_q) /T^4$
as a function of $\mu_I$ from the integral method, and different
temperatures. }
\label{fig:pressI}
\end{figure}

\begin{figure}[htb]
%\framebox[55mm]{\rule[-21mm]{0mm}{43mm}}
\psfig{file=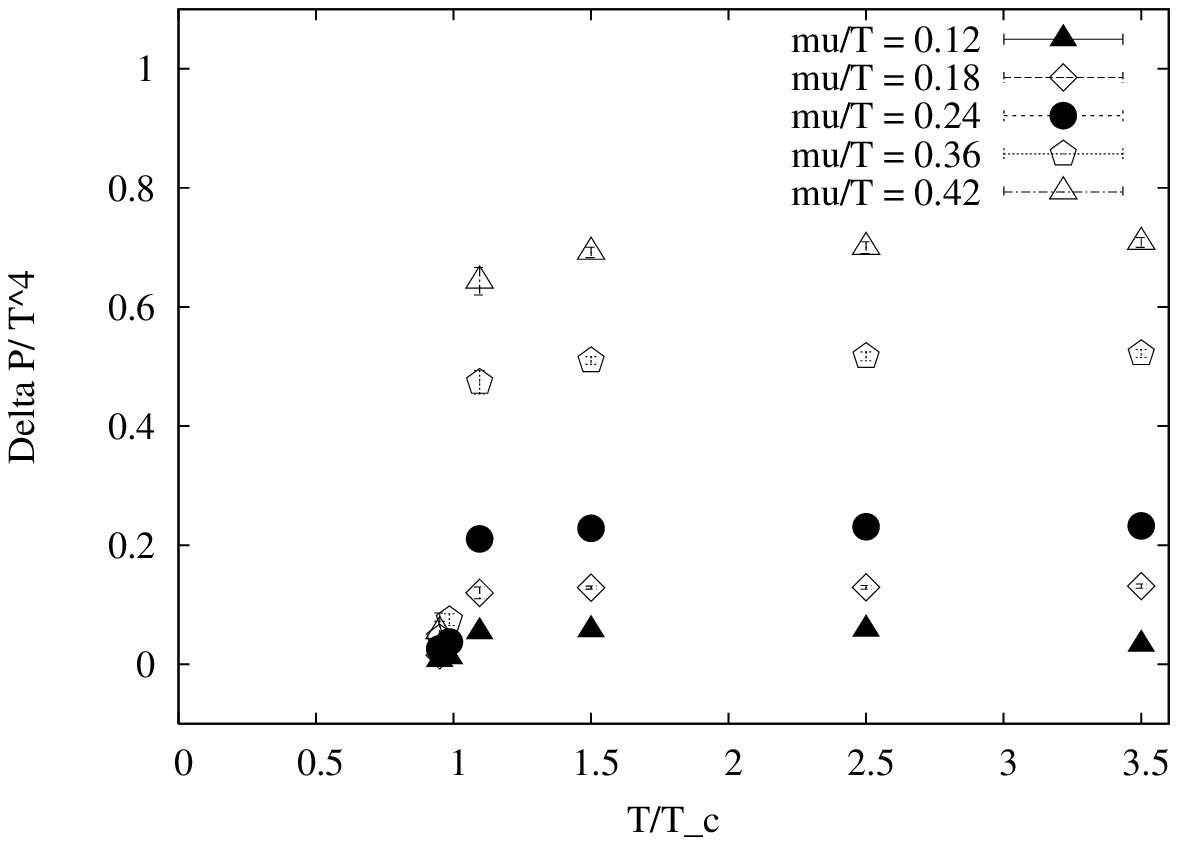,width=14 truecm}
\caption{Overview of the results for $\Delta P (T, \mu_I, m_q) /T^4$ 
as a function of $T/T_c$, and different values of $\mu_I/T$.}
\label{fig:pressII}
\end{figure}

In \cite{D'Elia:2003uy}, 
we  computed the fermionic contribution to the pressure
using the basic thermodynamic identity  
\begin{equation}
p(T,\mu,m_q) = \frac{T \partial ln {\cal Z}(V,T,\mu,m_q)}{\partial V }
\end{equation}
combined with  the naive tree level result for the Karsch coefficients 
\cite{karschcoe}.
By comparing the results for $\Delta P / T^4$ of \cite{D'Elia:2003uy}  
to our present results from the integral method, 
we note that the nonperturbative Karsch factors should be fairly large:
indeed, the tree level results exceed the numerical results from
the integral method; even ignoring the gluonic contribution, we see that
the correcting factor should be $\simeq .5$.

One further observable we shall consider 
is the derivative of the chiral condensate
with respect to the chemical potential.
This can be computed by numerical differentiation and 
will be used to estimate the mass dependence of the number
density according to 
the Maxwell relation \cite{Kogut:1983ia}
\begin{equation}
\frac{\partial \langle \bar \psi \psi \rangle (T,\mu,m_q) }{\partial \mu} = 
\frac{\partial n(T,\mu,m_q)}{ \partial m_q}
\end{equation}

\section{The Hadronic Phase and the Hadron Resonance Gas Model}

In this region observables are a continuous and periodic function
of $\mu_I/T$ , analytic continuation
in the $\mu^2 > 0$ half plane is always possible, but
interesting only when \\ $\chi_q(T, \mu=0) > 0$.

The analytic continuation of any 
observable $O$ is valid within the analyticity domain, i.e.
till $\mu < \mu_c(T)$, where $\mu_c(T)$ has to be measured independently.
The value of the analytic continuation of $O$ at
$\mu_c$, $O(\mu_c)$, defines the discontinuity
at the critical point, or, equivalently, its critical
value. This allows the identification of the order of the phase transition:
first, when $O(\mu_c) \ne 0$, second, when $O(\mu_c) = 0$.

Taylor expansion and Fourier decomposition are among the
natural parameterizations for the observables\cite{D'Elia:2002gd}.
In particular, the  analysis of the phase diagram in the temperature-imaginary 
chemical potential plane suggests
to use Fourier analysis in this region, as observables are periodic
and continuous there. Note that for
the simple one dimensional QCD model, which can be analytically solved
and is related to the partition function of QCD in the infinite
coupling limit,  
all of the Fourier coefficients but the first ones will be
zero.

For observables which are even ($O_e$) or odd ($O_o$) under
$\mu \to -\mu$    the Fourier series read:

\begin{eqnarray}
O_e(\mu_I, N_t) & = & \sum_n  a_{Fe}^{(n)} \cos (n N_t N_c \mu_I) \\
O_o(\mu_I, N_t) & = & \sum_n  a_{Fo}^{(n)} \sin (n N_t N_c \mu_I)
\end{eqnarray}
which is easily continued to real chemical potential:
\begin{eqnarray}
O_e(\mu_I, N_t) &  = & \sum_n  a_{Fe}^{(n)} \cosh (n N_t N_c \mu_I) \\
O_o(\mu_I, N_t) & = & \sum_n  a_{Fo}^{(n)} \sinh (n N_t N_c \mu_I) \, .
\end{eqnarray}

In our past Fourier analysis of the chiral condensate - obviously
an even observable -  
we limited ourselves to $n=0,1,2$ and we assessed the validity
of the fits via both the value of the $\chi^2/{\rm d.o.f.}$ and the stability
of  $a_{Fe}^{(0)}$ and $a_{Fe}^{(1)}$ given by one and two cosine fits:
we found that one cosine fit is actually enough to describe
the data up to $T \simeq T_c$ in the four flavor
model \cite{D'Elia:2002gd}; adding a
term  $\cos (2 N_t N_c \mu) = \cos (24 \mu)$ in the expansion 
did not modify the value of
the first coefficients and does not particularly  
improve the $\chi^2/{\rm d.o.f.}$. 

We present here the analogous analysis for the number density.

\begin{figure}[htb]
%\framebox[55mm]{\rule[-21mm]{0mm}{43mm}}
\psfig{file=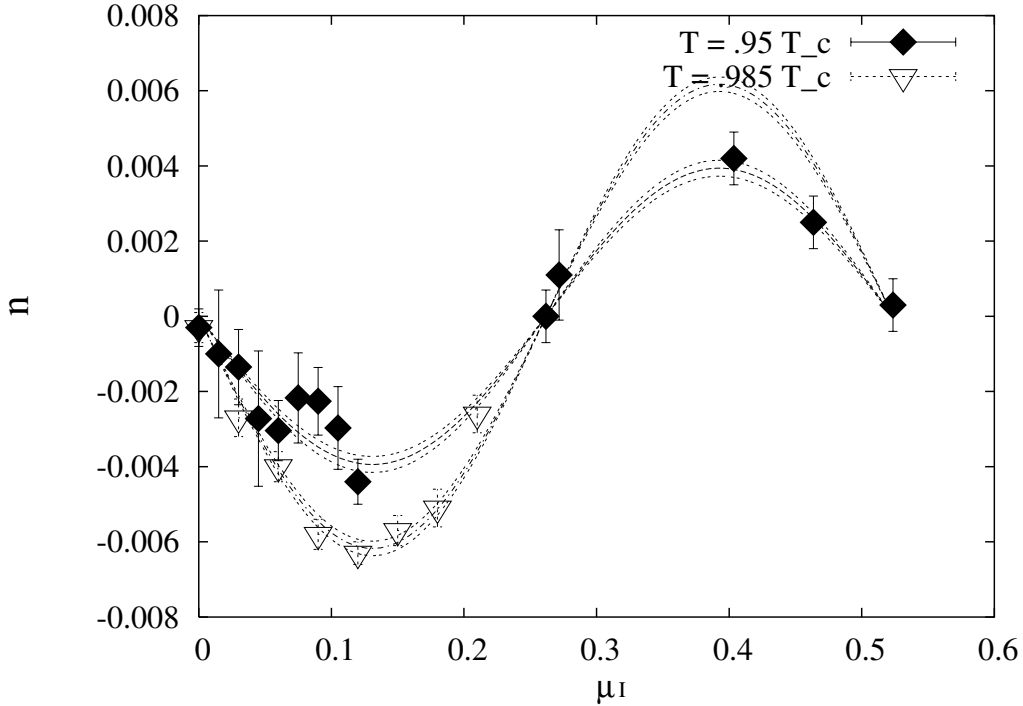,width=14 truecm}
\caption{One Fourier coefficient fit to the particle number in the hadronic
phase.}
\label{fig:fit_hadronic}
\end{figure}

A first round of fits was performed by setting all of the Fourier
coefficients but the first one $a_{Fo}^{(1)} = a_F$ to zero
\begin{equation}
n(T, \mu_I,m_q) = a_F sin (3 \mu_I / T)
\end{equation}
obtaining
\begin{eqnarray}
n(T,\mu_I,m_q) &=&  0.0039(2)sin(3 \mu_I/T)\\
n(T,\mu_I,m_q) &=&  0.0062(2)sin(3 \mu_I/T)
\end{eqnarray}
for $\beta = 5.010$ (upper, $\chi^2/{\rm d.o.f.} = 0.39$) 
and $\beta = 5.030$ ( $\chi^2/{\rm d.o.f.} = 1.06 $), or
equivalently $T \simeq  0.95 T_c$ and $ T \simeq 0.985 T_c$.
The results, and the relative errorbands, are shown in 
Figure \ref{fig:fit_hadronic}. 

To further assess the validity of this simple parametrization,
 we have also performed polynomial fits  of the form
\begin{equation}
f(x) = 12a_P^{(1)}x - 288a_P^{(3)}x^3
\end{equation}
The constants are chosen so that $a_P^{(1)} = a_F = a_P^{(3)}$ when
a simple sine fit is adequate.
At $\beta = 5.010$ we find $a_P^{(1)}= -0.0044(7)$ and 
$a_P^{(3)} = -0.009(4)$  ($\chi^2/{\rm d.o.f.} = 0.28 $), 
consistent with $a_P^{(1)} = a_F = a_P^{(3)}$, within
the large errors. At $\beta = 5.030$ we find $a_P^{(1)} = -0.0060(2)$
and $a_P^{(3)} = -0.0048(3)$ ($\chi^2/{\rm d.o.f.} = 0.69 $), 
again indicating that the the contribution
from a second Fourier coefficient is small, although perhaps
not negligible.

The obvious analytic continuation, representing the results for the
number density for these parameters from one
parameter Fourier fit, read:
\begin{eqnarray}
n(\beta = 5.010 (T \simeq 0.95 T_c) ,\mu,m_q) &=&  0.0039(2)sinh(3 \mu/T)\\
n(\beta = 5.030 (T \simeq 0.985 T_c) ,\mu,m_q) &=&  0.0062(2)sinh(3 \mu/T)
\end{eqnarray} 
A second order term in the Fourier series improves slightly the quality
of the fit but is poorly determined. It is however important to
determine the uncertainty on the analytic continuation and 
on the estimate of the critical density induced by this contribution:
indeed, it might well be that a term which is subleading in the
imaginary chemical potential domain becomes leading when the
results are analytically continued to the real plane.

\begin{figure}[htb]
%\framebox[55mm]{\rule[-21mm]{0mm}{43mm}}
\psfig{file=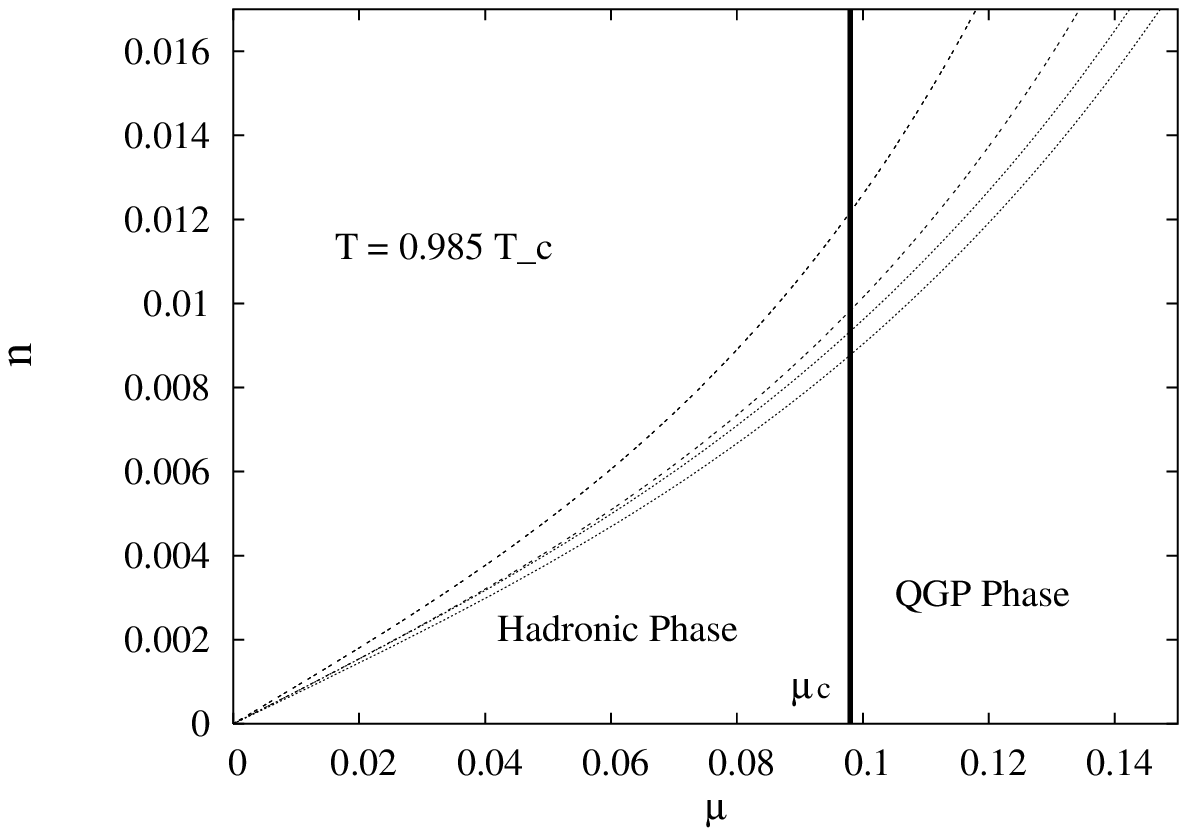,width=14 truecm}
\caption{The number density as a function of $\mu$ from analytic continuation;
the errors from one and two Fourier coefficient fits are shown. 
The vertical line marks $\mu_c$  from Ref. \cite{D'Elia:2002gd}, 
and the intercepts 
with the number density defines the critical density.}
\label{fig:fit_hadronic_criti}
\end{figure}

The situation is better seen in the plot
Figure \ref{fig:fit_hadronic_criti}, where we plot the results for
the analytic continuation to real $\mu$ with the
errorbands from the fits, for one (dotted line) and 
two (dashed) Fourier coefficients fit, for $\beta = 5.030$. 
On the same plot we mark with
a vertical line at $\mu = \mu_c$ the limit of validity of the analytic
continuation, which coincides with the limit of the hadronic phase.

Following the discussion above,
at $\mu = \mu_c$ we read off the plot $n(\mu_c) =  0.0087(30)$ (lattice units).
If we consider the second Fourier's coefficients the induced uncertainty grows
big, and we would obtain $n(\mu_c) = 0.011(1)$. 
All in all, the critical density 
we estimate at $T = 0.985 T_c$ is $n(\mu)/T^3 \simeq 0.6$.

To study (at least semiquantitatively) the mass dependence of the
results, we
consider the Maxwell relation
\begin{equation}
\partial \pbp / \partial \mu  = \partial n(\mu) /\partial m \, .
\end{equation}
The results for the chiral condensate can thus be used to
estimate the mass dependence.

In an attempt to introduce as less prejudice as possible, we have first numerically
derived the results for the chiral condensate. The results, although very noisy,
are in agreement with the derivative of the fitting function for the chiral
condensate itself. So we use the latter for the subsequent discussion.

Consider the parametrization for the chiral condensate
\begin{eqnarray}
\langle \bar\psi \psi \rangle (T, \mu, m_q)  &=& a_C cosh (3 \mu N_T) + b_C 
\end{eqnarray}
which combined with
\begin{eqnarray}
n(T, \mu, m_q) &=& a_n sinh (3 \mu N_T)  
\end{eqnarray}
gives
\begin{equation}
\frac{n(\mu,m_q + \Delta m_q) - n(\mu, m_q)}{n(\mu,m_q)} =
3 N_t \Delta m \frac{a_C}{a_n} \, .
\end{equation}

Combining the present results form the number density with the
old ones \cite{D'Elia:2002gd} for the chiral condensate, we obtain
\begin{eqnarray}
\frac{n(\mu,m_q + \Delta m_q) - n(\mu, m_q)}{n(\mu,m_q)} 
&\simeq & 2.5 \times 3 N_t \Delta m_q \\
\frac{n(\mu,m_q + \Delta m_q) - n(\mu, m_q)}{n(\mu,m_q)} 
&\simeq &  4 \times 3 N_t \Delta m_q 
\end{eqnarray}
for $T = 0.95 T_c$ ($\beta = 5.010$) and $T = 0.985 T_c$
($\beta = 5.030$) respectively.

\begin{figure}[htb]
%\framebox[55mm]{\rule[-21mm]{0mm}{43mm}}
\psfig{file=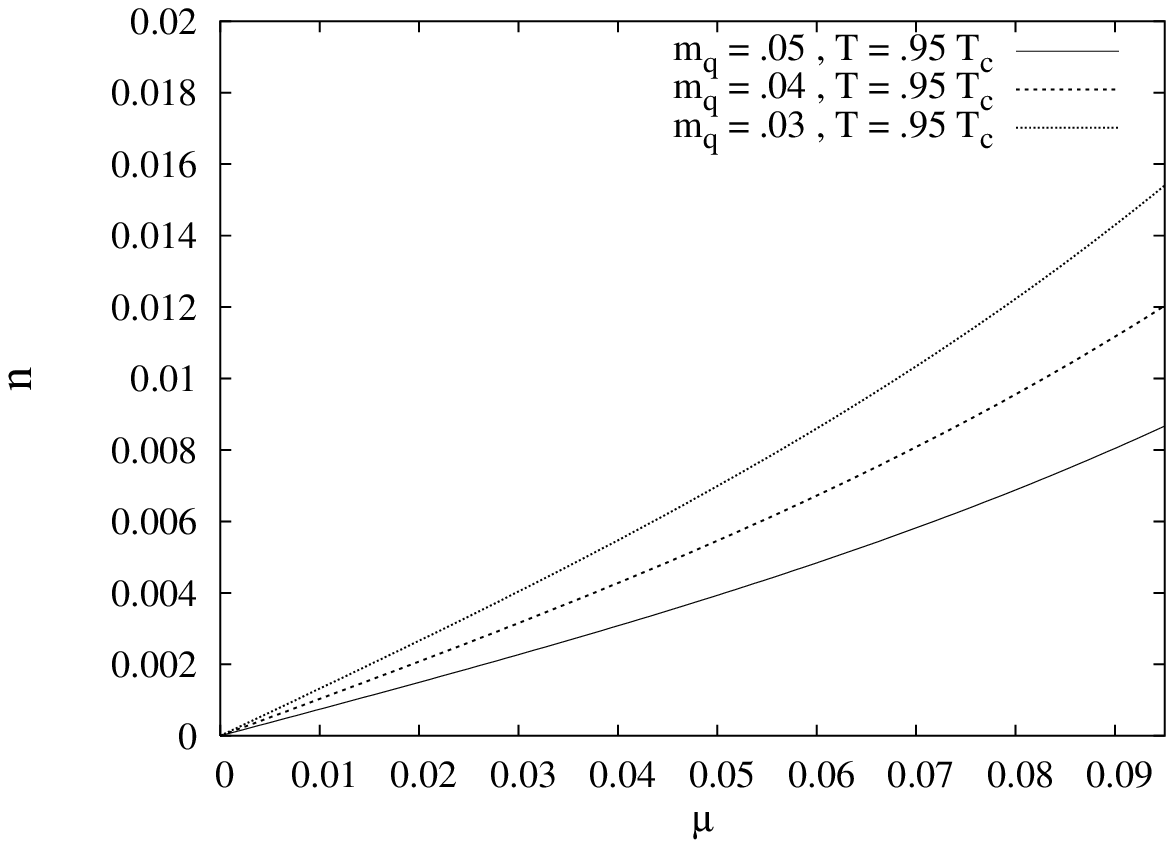,width=7 truecm}
%\end{figure}
%\begin{figure}[htb]
%\framebox[55mm]{\rule[-21mm]{0mm}{43mm}}
\psfig{file=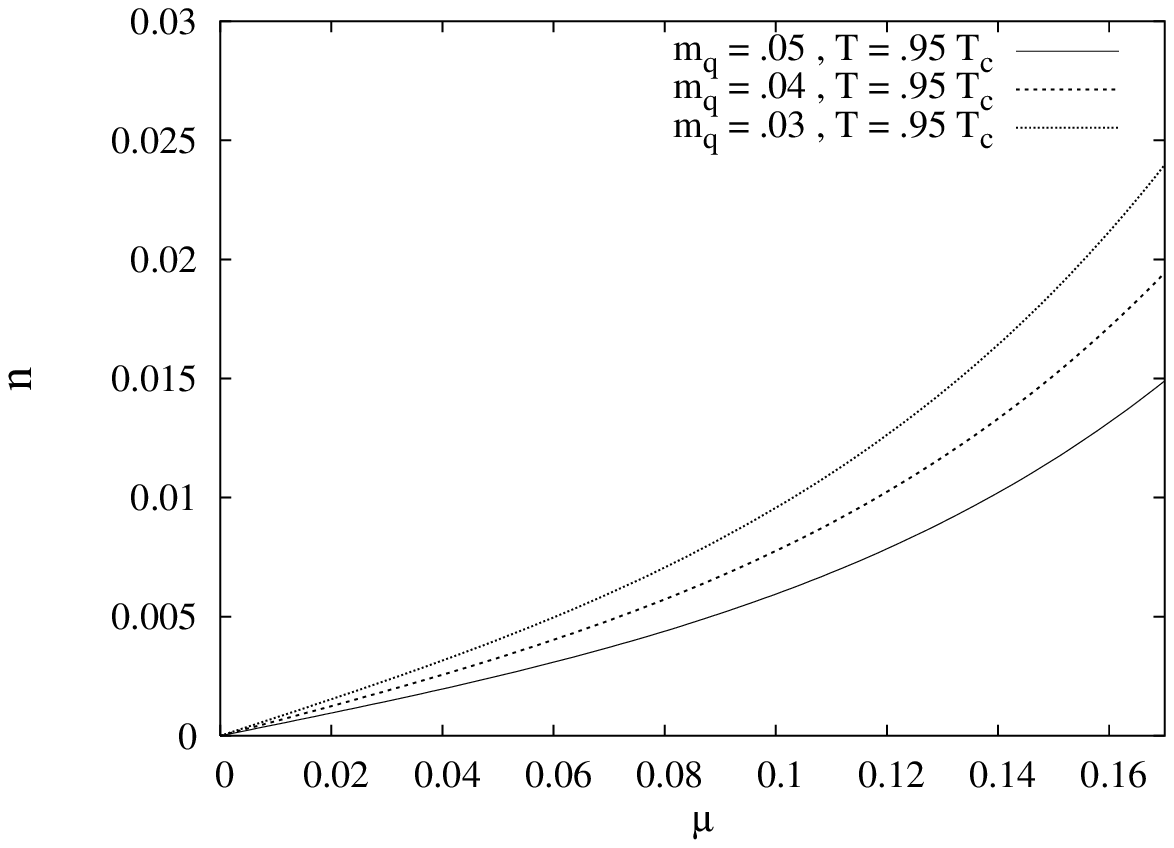,width=7 truecm}
\caption{The mass dependence of the number density at $T = .985 T_c$ 
(left) and $T = 0.95 T_c$ (right) }
\label{fig:mass_985_95}
\end{figure}

To illustrate this dependence, we  plot the numerical results
in Figure \ref{fig:mass_985_95}.
Note that in the plots $\mu$ range from $0.0$ to  $\mu_c(T)$: hence
the critical densities and their mass dependences can be read off at
the intercept with the righthand border of the plot.  

We meant here  to get an estimate 
of the mass dependence of the number density:
to this end we used for both temperatures only the results from
one Fourier coefficient fit,and we omit the errors 
(which, for $T = 0.985 T_c$, can be read off Figure \ref{fig:fit_hadronic}.)

To summaries our findings,
our previous results for the chiral condensate \cite{D'Elia:2002gd} 
and the present ones for the number density are consistent with 
$\Delta P \propto (cosh (\mu_B/T) - 1)$
in the broken phase. There is room for a small deviation especially at 
higher T  values.
 
In  \cite{Karsch:2003zq} it was shown that
the data obtained from an expanded reweighting behave in the same
way, and it was pointed out that this result is the
one expected from an hadron resonance gas model.

\section{The QGP phase and the Equation of State}

At high temperature, in the weak coupling regime,
perturbation theory might serve as a guidance, 
suggesting that the first few terms of the
Taylor expansion might be adequate in
a wider range of chemical potentials.  
So, at a variance with the expansion in the hadronic phase,
where the natural parametrization is given by a Fourier analysis,
in this phase the natural parametrization for the
grand partition function is a polynomial. 

The leading order result for the pressure$p(T,\mu)$ in the
massless limit is easily computed, given
that at zero coupling the massless theory reduces to a non--interacting gas
of quarks and gluons, yielding for the pressure
\begin{equation}
p(T, \mu) = \frac{\pi^2}{45} T^4 \left(8 + 7 N_c \frac{n_f}{4}\right) +  \frac{n_f}{2} \mu^2 T^2
+  \frac{n_f}{4 \pi^2} \mu^4 \, .
\end{equation}
Obviously, when analytically continued to the negative
$\mu^2$ side, this gives
\begin{equation}
p(T, \mu_I) = \frac{\pi^2}{45} T^4 \left(8 + 7 N_c \frac{n_f}{4}\right) -  \frac{n_f}{2} \mu_I^2 T^2
+  \frac{n_f}{4 \pi^2} \mu_I^4 \, .
\end{equation}
Because of the Roberge Weiss periodicity this polynomial
behavior should be cut at the Roberge Weiss transition 
$\mu_I = \pi T / 3$:
this is consistent with the Roberge Weiss critical line
being strongly first order at high temperature.
We discuss first the results of the fits of the
number density to polynomial form; then we contrast these results
with a free field behavior. 

The considerations above suggests 
a natural ansatz for the behavior of the number density in this
phase as a simple polynomial with only odd powers. 
We performed then fits to 
\begin{equation}
n(T,\mu_I) = a(T) \mu_I - b(T) \mu_I^3
\end{equation}
whose obvious analytic continuation is
\begin{equation}
n(T, \mu) = a(T) \mu + b(T) \mu^3 \, .
\end{equation}
Note again that $a(T) = \chi_q(T, \mu=0)$.

The results of the fits are given in Table \ref{polfitn}, upper rows.
To assess the relevance of the third order term we have performed
fits with $b(T)=0$, whose results are summarized in the bottom rows of
Table \ref{polfitn}.
As usual, the quality of the fit worsten slightly, while
the first coefficient $a(T)$ remains compatible with that estimated by
a two parameter fit.

On the other hand,  at $\mu_I = 0.1$ (for instance) 
the contribution of the third order term to the number density 
of the free lattice gas
is below two percent :  a fair set of measurements
with this precision around $\mu_I = 0.1$ would then be needed to
disentangle the third order term from the error,  and 
it comes as no surprise that
within the current precision is not possible to safely estimate it.

\begin{table}
\caption{Coefficients of a polynomial fit 
for the number density in the Quark Gluon Plasma phase.}
\begin{ruledtabular}
\begin{tabular}{l l l l }
$\beta$ & $a^P_{1}$ & $a^P_{3}$ & $\chi^2 /d.o.f.$ \\
\hline
5.10 &  -0.4646(68)  & 2.02 (60)  &  0.89 \\
5.310 & -0.4994(40) &  1.83(64)   &  0.92 \\
5.650 & -0.5129 (43) &   2.36(82)    & 1.65 \\
5.869 & -0.5087 (16) &   0.89(28)    & 0.30 \\
\hline
5.10 &  -0.4442(42)  & 0  &  1.66 \\
5.310 & -0.4897(29) &  0   &  1.87 \\
5.650 & -0.5026 (31) &  0    & 2.88 \\
5.869 & -0.5039 (7) &  0    & 0.58 \\
\end{tabular}
\label{polfitn}
\end{ruledtabular}
\end{table}

\begin{figure}[htb]
%\framebox[55mm]{\rule[-21mm]{0mm}{43mm}}
\psfig{file=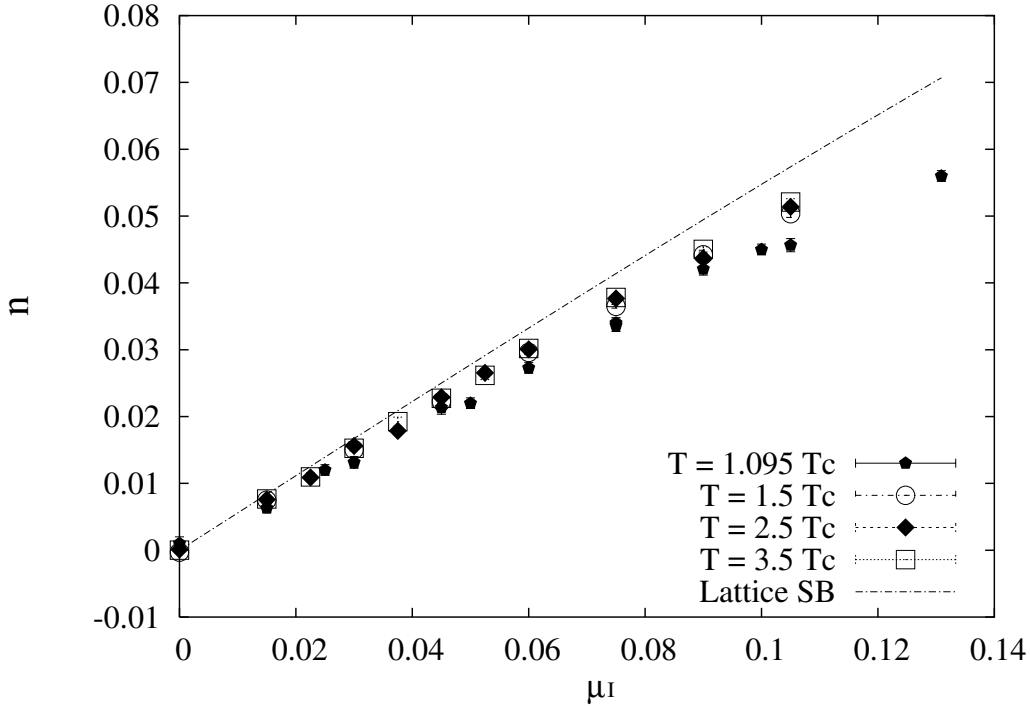,width=14 truecm}
\caption{Results for the particle number as a function of imaginary
chemical potential. The dotted line is the free field 
for imaginary chemical potential.}
\label{fig:number_hot}
\end{figure}

In Figure \ref{fig:number_hot} we show the results for the particle
number at $T=1.5 T_c$, $T=2.5 T_c$, $T=3.5 T_c$  
as a function of the imaginary chemical potential, together with 
the free lattice result (because of the known discrepancies between 
the lattice and continuum
behavior in the free case at $N_T = 4$, we used lattice free results
for this comparison, as was already done for the quark number susceptibility
in Figure 4 above).

Some deviations are apparent, whose origin we would like to
understand.  It would be however 
arduous, given the strong lattice artifacts, to try to make
contact with a rigorous perturbative analysis carried out in
the continuum \cite{Vuorinen:2004rd,Vuorinen:2003fs,Ipp:2003yz}.  
Rather then attempting that, we parametrize
the deviation  from a free field behavior as 
\cite{Szabo:2003kg,Letessier:2003uj}
\begin{equation}
\Delta P (T, \mu)  =  f(T, \mu) P^L_{free}(T, \mu) 
\end{equation}
where $P^L_{free}(T, \mu)$ is the lattice free result for the pressure.
For instance, in the discussion of Ref. \cite {Letessier:2003uj}
\begin{equation}
f(T, \mu) = 2(1 - 2 \alpha_s/ \pi)
\end{equation}
and the crucial point was that $\alpha_s$ is $\mu$ dependent.

We can search for such a non trivial prefactor $f(T, \mu)$ by taking 
the ratio between the numerical data and the lattice
free field result $ n^L_{free}(\mu_I)$  at imaginary chemical potential:
\begin{equation} 
R(T, \mu_I) = \frac{ n(T,  \mu_I)}{n^L_{free}( \mu_I)}
\end{equation}
A non-trivial (i.e.
not a constant) $R(T, \mu_I)$ would indicate a non-trivial 
$f(T, \mu)$.

\begin{figure}[htb]
%\framebox[55mm]{\rule[-21mm]{0mm}{43mm}}
\psfig{file=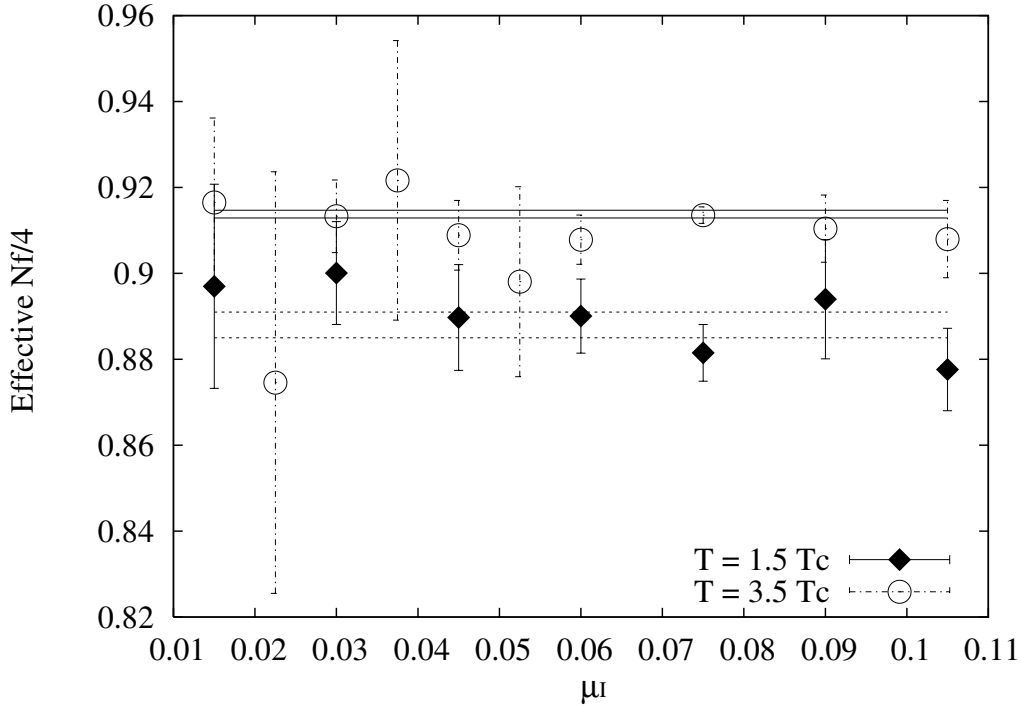,width=14 truecm}
\caption{Ratio of effective number of active flavors 
to the continuum $N_f=4$ as estimated from the ratio
of the lattice results to the lattice free field.}
\label{fig:effective_nf_hot}
\end{figure}

In Fig. \ref{fig:effective_nf_hot}   we plot $R(T, \mu_I)$ 
versus $\mu_I/T$: we see that $R(T, \mu_i)$ is constant within
errors, so that our data do not permit to distinguish a non trivial
factor within the error bars: rather, 
the results for $T \ge 1.5 T_c$ seem consistent with a free lattice
gas, with an fixed effective number of flavors $N^{eff}_f(T)/ 4 =  R(T) $:
$N^{eff}_f=  0.92 \times 4$ for $T=3.5 T_c$,  and 
$N^{eff}_f = 0.89 \times 4$ for $T = 1.5 T_c$.
These results confirm those obtained at $\mu=0$ from the quark number
susceptibility in Figure 4 
and extend them over a finite range of chemical potentials

One last remark concerns the mass dependence of the results, which,
as in the broken phase, can be computed from the derivative of the
chiral condensate. In the chiral limit this gives 
$\frac {\partial n}{\partial m} = 0$ , since the chiral condensate
is identically zero. We have verified that 
$\frac {\partial n}{\partial m}$ remains very small compared to
$n$ itself: in a nutshell, in the quark gluon plasma phase $<\bar \psi \psi>$
is very small (zero in the chiral limit), while the number density grows
larger, and this implies that the mass sensitivity is greatly reduced
with respect to that in the broken phase.

\begin{figure}[htb]
%\framebox[55mm]{\rule[-21mm]{0mm}{43mm}}
\psfig{file=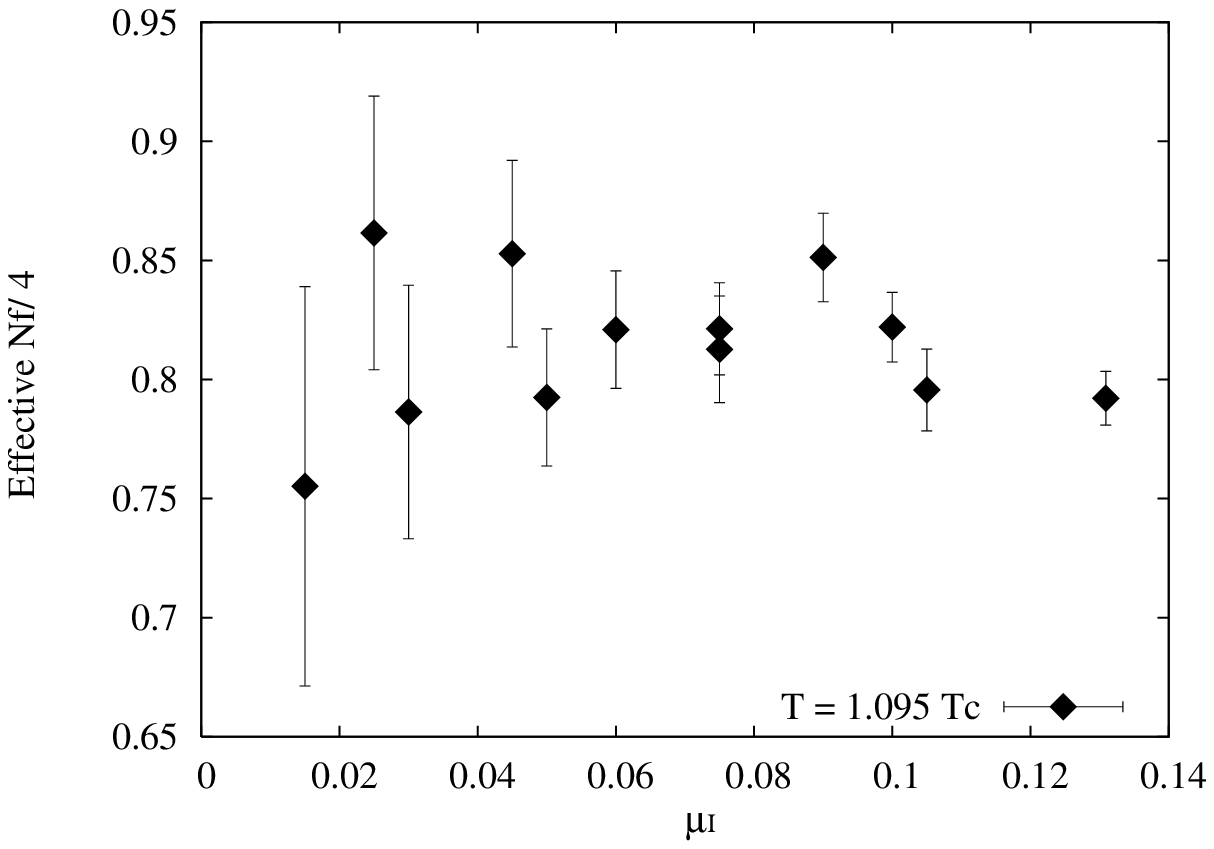,width=14 truecm}
\caption{Deviation from the free field behavior in the RW regime }
\label{fig:effective_NF_RW}
\end{figure}

\section{The intermediate regime $T_c < T < T_E$}

The discussions presented above bring  very naturally to the consideration
of a dynamical region which is comprised between the deconfinement transition,
and the endpoint of the Roberge Weiss transition.

In this dynamical region
the analytic continuation is valid till $\mu = \infty$ 
but the interval accessible to the simulations at
imaginary $\mu$ is small, as
simulations in this area hits the chiral critical
line for $\mu^2 < 0$.

In Figure \ref{fig:effective_NF_RW} we repeat the same analysis
for the number density done in the previous Section,
but for $T/T_c = 1.095$.
We see that the results are noisier than those as higher temperature,
and it is difficult to draw firm conclusions. Anyway,
they might still accommodate some deviation from a simple free field
with a reduced effective number of flavor.

Let us make some general consideration about the 
thermodynamic behavior in this region 
by considering the critical line at imaginary
chemical potential, Let us consider first the case of a second
order transition: the analytic continuation
of the polynomial predicted 
by perturbation theory for positive $\mu^2$ 
would hardly reproduce the correct critical behavior
at the second order phase transition for $\mu^2 < 0$.
In fact, for a second order
chiral transition at negative $\mu^2$, 
$\Delta P (T, \mu^2) \propto (\mu^2 - {\mu^2}_c)^\chi $, 
where $\chi$ is a generic exponent. As the window between the critical
line  and the $\mu=0$ axis is anyway small, such behavior - possibly
with subcritical corrections - should persist in the proximity of the
real axis.
For generic values of the exponent a second
order chiral transition seems incompatible with a free field behavior.
The same discussion  can be repeated for a first order transition of 
finite strength, by trading the critical point $\mu_c$ with 
the spinodal point $\mu^*$ . So deviations 
from free field are to be expected in this intermediate regime.

\section{Summary and Outlook}

We have gained a good understanding 
of the strength and weakness of the method. First,
the imaginary chemical potential approach is not limited
to small volumes  
(aside from the usual limitations of any lattice calculations).
Next,  physical observables can be directly computed by usual methods,
and their analytic continuation, or, extrapolation, can be pushed
up to the critical line, thus providing estimates of the critical
values and discontinuities. In addition to that, 
the method  provides a natural test bed for analytic models or 
calculations, which can be analytically
continued to imaginary chemical potential, and directly contrasted
with the numerical results. On the weak side,  corrections
which are subleading for imaginary chemical potential, or
for $\mu^2 < 0$,  might become
leading in the real domain, $\mu^2 \ge 0.$: it is important
to try and cross check different analytic parameterizations,
and we have given a few examples of this procedure in our 
analysis.

We have obtained results on the four flavor model   
for $.985 T_c < T < 3.5 T_c$, and $\mu_B \le 500 MeV$.

Concerning the critical line,  we have 
studied in detail the chiral and ``deconfining'' transition 
at a selected value of $\mu_I$ and confirmed that they 
remain correlated, showing also the complete correlation between
the Monte Carlo time histories of the Polyakov loop and the chiral condensate
around the phase transition. As explained in Ref. \cite{D'Elia:2002gd} 
and in Section II above,
this, together with the observation
of their correlation at zero chemical potential, implies the
equality of the critical temperature
\begin{equation}
 T^{c}_c(\mu) = T^{d}_c(\mu) 
\end{equation}
also for real chemical potential.

In the hadronic phase the corrections to 
$n(T, \mu_B) \propto  sinh (\mu_B/T)$ are very small. This confirms and
completes the finding of Ref. \cite{D'Elia:2002gd} 
where we did show that the chiral
condensate behaves as
$\pbp (T, \mu_B) \propto  cosh (\mu_B/T) + c$.
In conclusion our results in the hadronic
phase are consistent with an hadron resonance gas model,
possibly with small corrections close to $T_c$. 
Again in the hadronic phase
we have calculated the baryon density in the hadronic phase,
and estimated its critical value 
$n(\mu_c, T = .985 T_c, m_q = .05)/T_c^3 \simeq 0.6$; the mass dependence
has been inferred  from the Maxwell relation giving
$\Delta n = - 4.0 3 \Delta m_q / T$.

In the high temperature regime, for $T\ge 1.5 T_c$ the results are
compatible with lattice Stefan-Boltzmann with
an effective fixed  number of active flavors $\simeq 0.92 \times 4$ 
for T=3.5 $T_c$ 
and $\simeq 0.89 \times 4 $ for $T = 2.5 T_c$. 
We found that the mass dependence is very small
in this region.

We discussed the interplay between thermodynamics and
chiral transition in the region comprised between the critical point
$T_c$ and the endpoint of the Roberge Weiss transition 
$T_E$. We noted the possibility of non-trivial deviations
from a free lattice field, possibly 
connected with the chiral transition at $\mu^2 < 0$.

As for future applications, the method seems ideally suited for more
detailed comparisons with analytic models, and , more important,
nothing prevents its extension to larger lattices.

We think that the performance could be further improved by considering
hybrid methods which combines the imaginary chemical potential approach
with other methods, 
for instance by making use of reweighting 
\cite{Fodor:2002km,Crompton:2001ws}  or direct
calculations of derivatives \cite{Gavai:2003mf} 
at nonzero $\mu$ to improve the accuracy of the results at negative $\mu^2$.

Finally, the study of discontinuities, as sketched in Sect. IV above,
might offer an alternative approach to the study
of endpoints and tricritical points.

\section*{Acknowledgments}

We would like to thank F. Csikor, Ph. de Forcrand, R. Gavai, S. Gupta 
and A. Vuorinen for helpful discussions. In addition, 
MPL wishes to thank the Institute for Nuclear Theory at the University
of Washington for its hospitality and the Department of Energy
for partial support during the completion of this work.
This work has been partially supported by MIUR. The simulations
were performed on  the APEmille computer of 
{\em Consorzio Ricerca del Gran Sasso}: we wish to thank Enrico Bellotti,
Aurelio Grillo  and in particular Giuseppe Di Carlo 
for providing access to this facility as well as for their kind help.

\end{document}